\documentclass{ws-rv9x6}
\usepackage{subfigure}   
\usepackage{ws-rv-thm}   
\usepackage{ws-rv-van}   
\makeindex

\usepackage{hyperref}

\begin{document}

\chapter[Forensic Data Analytics for Anomaly Detection in Evolving Networks]{Forensic Data Analytics for Anomaly Detection in Evolving Networks\label{ra_ch1}}

\author[L. Yang et al.]{Li Yang$^*$, Abdallah Moubayed$^*$,  Abdallah Shami$^*$, Amine Boukhtouta$^\dagger$, Parisa Heidari$^\ddagger$, Stere Preda$^\dagger$, Richard Brunner$^\P$, Daniel Migault$^\dagger$, and Adel Larabi$^\dagger$
}

\address{$^*$Western University, London, Ontario, Canada; e-mails:\{lyang339, amoubaye, abdallah.shami\}@uwo.ca\\
$^\dagger$Ericsson, Montreal, Quebec, Canada; e-mails:\{amine.boukhtouta, stere.preda, daniel.migault, adel.larabi\}@ericsson.com\\
$^\ddagger$IBM, Montreal, Quebec, Canada; e-mail: parisa.heidari@ibm.com\\
$^\P$Log5Data, Montreal, Quebec, Canada; e-mail: rick@log5data
}

\begin{abstract}
In the prevailing convergence of traditional infrastructure-based deployment (\textit{i.e.}, Telco and industry operational networks) towards evolving deployments enabled by 5G and virtualization, there is a keen interest in elaborating effective security controls to protect these deployments in-depth. By considering key enabling technologies like 5G and virtualization, evolving networks are democratized, facilitating the establishment of point presences integrating different business models ranging from media, dynamic web content, gaming, and a plethora of IoT use cases. Despite the increasing services provided by evolving networks, many cybercrimes and attacks have been launched in evolving networks to perform malicious activities. Due to the limitations of traditional security artifacts (\textit{e.g.}, firewalls and intrusion detection systems), the research on digital forensic data analytics has attracted more attention. Digital forensic analytics enables people to derive detailed information and comprehensive conclusions from different perspectives of cybercrimes to assist in convicting criminals and preventing future crimes. This chapter presents a digital analytics framework for network anomaly detection, including multi-perspective feature engineering, unsupervised anomaly detection, and comprehensive result correction procedures. Experiments on real-world evolving network data show the effectiveness of the proposed forensic data analytics solution.
\end{abstract}


\body

\section{Introduction}
The evolution of modern networks enables the shift from traditional infrastructure towards key enabling technologies, like virtualization, cloud, and 5G \cite{kodali2020a}. These technologies are meant to settle the integration of fully-fledged business models in different points of presence, including content delivery, gaming, Internet of Things (IoT), etc. The Ericsson June 2020 Mobility Report \cite{ericcson2020a} highlights the rapid growth of 5G networks, infrastructure, applications, and end-user services. Due to the COVID-19 pandemic, 88\% of professionals use online video calls in their work and personal life. Communication Service Providers (CSPs) face the challenge of delivering resilient and secure networks, as well as innovative service offerings \cite{ericcson2020a}. Ericsson sees Intelligent security management as a business accelerator. CSPs need support in automating, scaling, and adapting their security posture to stay protected against threats in this evolving 5G network landscape.

With the development of evolving networks, various types of network devices and entities have generated large volumes of data and traces \cite{hou2020a}. During network communications, massive data will be collected at the network level. Client-centric forensic data is a major source of network data, including history logs, access logs, chat logs, cookies, system logs, etc. \cite{hou2020a}. The extensive digital traces in network environments can potentially give insights into the actions and behaviors of network users and devices.

Most research on network data and forensic analytics has focused on static networks, in which single observations or combined historical datasets that do not change over time are collected and used \cite{kodali2020a}. However, the majority of modern networks are connected by evolving systems with dynamic activities, named evolving networks \cite{kodali2020a}. In real-world applications, modern networks evolve abruptly and continuously. For example, real-world network devices and services usually face continuous upgrades due to new user requirements, causing the networks to evolve over time. Cyber-attacks will also result in dramatic changes in communication networks.

Thus, it is valuable to conduct research in detecting abnormal changes in evolving networks for modern network protection. Accurate anomaly detection is crucial for decision-making and the timely execution of necessary actions. Additionally, sudden changes often occur only in certain nodes of a large evolving network \cite{kodali2020a}. Hence, it is important to locate the affected subnetworks and nodes, which can help us determine the root cause of anomalous events and take proper countermeasures. 

Network anomalies can be classified into two primary types: legitimate anomalies and network attacks \cite{thottan2003a}. Legitimate anomalies are network issues relating to operational failures and performance, such as server failures, misconfigurations, crowd events, etc. Network attacks and cybercrimes are anomalies that are launched by malicious attackers and often cause severe consequences. There have been a growing number of cybercrimes and attacks involving the devices and services of evolving networks. Not only may network devices be targets of cyber-attacks, but they can also be exploited to launch attacks \cite{hou2020a}. Cybercrimes and attacks often result in harmful consequences, such as service unavailability, network congestion, financial losses, and other severe issues \cite{deng2008a}. 

In this work, we focus on the cyber-attacks that target network services, including Denial of Service (DoS) attacks and Cache Pollution Attacks (CPAs), so-called \textit{service targeting attacks}. Since service targeting attacks are capable of disrupting services and compromising devices of evolving networks, it is essential to develop effective forensics approaches to identify malicious behaviors and cyber-attacks \cite{koroniotis2019a}. Network forensic analytics techniques enable us to identify the attackers and compromised networks/nodes, so as to take proper countermeasures to defend against attacks and address the network issues.

With the continuously growing volumes of digital data and the development of high technology crimes, the expense of conducting cybercrime investigations has been increasing dramatically \cite{kao2018a}. Thus, the research on digital forensic analytics has attracted significant attention, especially in dealing with large datasets. Digital forensics plays a vital role in crime reconstruction and evidence generation in court. Thus, it is crucial to develop digital forensic analytics techniques to discover more useful information about cyber-attacks, such as the attackers, victims, compromised devices, places, time periods, and means of attack. Network data analytics and forensic analytics have become essential techniques in evolving networks, and are expected to become more critical in the near future. 

Digital forensics is the process of discovering and evaluating relevant details about an event of interest to get a greater comprehension of the event \cite{hou2020a}. The event traces on the digital system are used by forensic investigators to expose the truth of the event. According to the National Institute of Standards and Technology (NIST) recommendation \cite{kent2006a}, digital forensics is also defined as the process of collecting, examining, analyzing, identifying, and presenting (CEAIP) digital data, aiming to transform the original information into digital evidence and provide investigation results using forensic methods  \cite{koroniotis2019a}. 

The forensic analytics techniques should be applied to network data to derive conclusions about the occurred cybercrimes, such as the attacker devices, their motivations or objectives, their attack methods or types, their exploited services, and the compromised devices. By connecting the forensic process to the data analytics process, Data Mining (DM) and Machine Learning (ML) algorithms can be utilized to construct network security models capable of effectively identifying cybercrimes. DM and ML techniques are able to discover useful information and trends that are difficult to be found by human observation. Thus, ML algorithms can be integrated with forensic analytics methods to handle the ever-increasing volume of data and information \cite{quick2014a}. ML algorithms are able to detect numerically abnormal patterns, while forensic analytics can be used to determine the real attack patterns and identify the relevant crime information, like affected devices and malicious attacker Internet Protocol (IP) addresses. 

A major challenge in digital forensics is to locate relevant information in large volumes of data \cite{quick2014a}. Additionally, many traditional anomaly detection solutions (\textit{e.g.}, ensemble-based Anomaly Detection System (ADS) \cite{moubayed2021b}, multi-stage ADS \cite{injadat2020a}, tree-based ADS \cite{yang2019a}, and Convolution Neural Network (CNN)-based ADS\cite{yang2022a})  only focus on the identification process through ML model learning, which is a small subset of the large digital forensic analytics process. Thus, this chapter presents a comprehensive unsupervised anomaly detection framework applicable to the forensic analytics of a wide range of information and evidence. 

On the other hand, for network data analytics, the main difference between static and evolving network analytics is that static network analysis uses single observations of the network without time correlations, while evolving network analysis extracts evolving features of the network by retaining time-series information \cite{wu2018a}. Hence, this chapter also proposes a multi-perspective feature engineering method to extract time-based features and retain time-series correlations for effective evolving network data analytics. 

The proposed multi-perspective feature engineering approach and the comprehensive unsupervised anomaly detection framework aim to fingerprint malicious clients or IoTs in the IP space and abnormal content. In addition, it allows the identification of targeted service nodes. The research and validation of the unsupervised anomaly detection framework are based on real-world service data logs collected from dynamic evolving networks. The large-sized service data consists of more than 452 million unlabeled lines of logs. Through experiments, the abnormal network entities (\textit{i.e.}, malicious IPs, abnormal contents, and compromised nodes) and their corresponding attack types are detected effectively using the proposed multi-perspective anomaly detection framework.

This chapter is organized as follows. Section \ref{back} presents the background of digital forensic analytics and service targeting attacks. Section \ref{lit} provides the literature review of forensic analytics and network anomaly detection. Section \ref{prop} presents an overview of the proposed solution. Sections \ref{pre}, \ref{unsu}, and \ref{vali} discuss each major procedure of the proposed solution in detail, including data pre-processing and feature engineering, unsupervised anomaly detection, and result correction. Section \ref{conc} concludes this chapter.

\section{Background}
\label{back}
This section provides backgrounds on the service targeting attacks in evolving networks and presents the digital forensic analytics procedures that can be used for network anomaly detection.

\subsection{Service Targeting Attacks in Evolving Networks}
\label{STA}
By service targeting attacks, we mean any type of disturbances performed to impact the functioning of serving nodes in evolving networks. By serving nodes, we mean any masterpiece used to integrate different business models like message brokers for IoTs, gateways, proxies, caching nodes, etc. \cite{moubayed2021a}.

The goal of cyber-attackers is usually either taking down a service or altering its inner workings \cite{abdallah2015a}. The former, commonly known as DoS attacks \cite{deng2008a}, is meant to overwhelm serving nodes and reduce their capability to drive well-functioning of services, resulting in sustainability impact. The latter is meant to take advantage of misconfigurations or vulnerabilities of serving mechanisms for mid or long-term impact on the sustainability of service without taking it down, resulting in stealthy attacks. 

For the sake of illustration, in the context of clouds, network threats can take down serving nodes resulting in a denial of service or misconfiguring internal parameters to impact serving nodes' production on-premises resulting in a sustainability impact. In the context of content delivery networks, CPAs \cite{deng2008a} can be launched by polluting the cache space with many unpopular or illegitimate contents; therefore, legitimate clients will get many cache misses for popular files, making the cache mechanism less effective resulting in a sustainability impact. An insider threat can tweak caching parameters resulting in impacting the effectiveness of the caching mechanism. This chapter focuses on the detection of service targeting attacks and the related network entities to present digital evidence through digital forensic analytics methods.

\subsection{Digital Forensic Analytics}
\label{DFA}
The standard Digital Forensic Analytics (DFA) process can be divided into five consecutive steps: collection, examination, identification, analysis, and presentation \cite{kao2018a}. Their definitions and challenges in evolving network analytics are as follows \cite{kent2006a}:
\begin{enumerate}
\item \textit{Collection}: The collection process aims to collect and integrate data from digital devices for subsequent procedures of digital forensics. Error data and uncertain data should be avoided in the data collection process. In evolving networks, network data is often acquired from a variety of nodes or other data sources with different attributes or formats. Additionally, network data sources often generate data at different speeds, making it challenging to integrate network data. Thus, it is critical to use proper data collection and integration methods in order to obtain high-quality data for further analysis.
\item \textit{Examination}: During the examination process, the collected data is examined and prepared for further identification and analysis. As original datasets are often collected from different sources, a preliminary data evaluation and reconstruction should be conducted to extract and interpret important digital evidence. Due to the limitations of raw data, there are many challenges associated with the examination process of evolving network data analytics. First, raw data often includes a large number of missing and error samples that should be eliminated or imputed in order to prevent noisy data from impairing the analytics results. Second, different class distributions (\textit{i.e.}, unbalanced data) and feature ranges have a negative effect on the performance of analytics models. They can be addressed by using appropriate data sampling and normalization approaches. Third, the raw network data is usually not the optimal data with the most appropriate attributes/features. Thus, proper feature engineering should be conducted to extract and select the most relevant network features for cyber-attack detection.
Many digital forensic technologies, like Automated Machine Learning (AutoML) techniques \cite{he2021a,yang2020a}, have been developed to automatically conduct the examination and pre-processing procedures to reduce human efforts and potential mistakes. 
\item \textit{Identification}: The identification process aims to identify the relevant digital objects, including the events, people, items, and methods related to the case, based on the analytics of the examined data. In the anomaly detection problems in the context of evolving networks, ML and DM algorithms are widely-used to develop classifiers that can identify the benign and abnormal events or cyber-attacks based on their patterns. 
As the ML-based identification approaches (\textit{e.g.}, Conti \textit{et al.} \cite{conti2013a} and Yang \textit{et al.} \cite{yang2022b}) usually suffer from high-false positive rates, further analysis on the identification results is often required.
\item \textit{Analysis}: In the analysis procedure, the digital objects obtained in the identification procedure are analyzed according to their relevance and reliability. The digital objects that meet the requirements will be chosen as digital evidence and investigation results. In the anomaly detection problems in the context of evolving networks, analysis can be conducted to validate and correct the identification results obtained by ML and DM models. For accurate analysis, human efforts and expert knowledge are often required to obtain sufficient and valid evidence. 
\item \textit{Presentation}: The presentation process aims to conclude the investigation results and present them in a proper way, so that the relevant staff and police officers can understand the investigation results clearly and use them for criminal conviction and accountability. In the anomaly detection problems in the context of evolving networks, as the investigation results often involve machine-readable data or technical terms, the main difficulty is to make the results user-friendly and human-understandable. Moreover, proper data visualization methods are also helpful for the presentation of network anomaly detection results.
\end{enumerate}

The proposed solution is designed based on the typical DFA process described above, which will be presented in Section \ref{overview}.

\section{Literature Review}
\label{lit}
In this section, a review on the existing literature is conducted, which includes general network anomaly detection methods, digital forensic data analytics techniques, service targeting attack detection approaches, and cybercrime-related entity detection methods.
\subsection{Network Anomaly Detection}
Network anomaly detection refers to the process of analyzing network packets to identify abnormal network events and behaviors for network security enhancement. This subsection focuses on the network anomaly detection systems that aim to identify malicious cyber-attacks and threats in modern networks by network traffic data analytics. Yang \textit{et al.} \cite{yang2022a} propose a CNN and transfer learning-based intrusion detection system for the Internet of Vehicles. The proposed system can transform general network traffic data into images to better fingerprint the patterns of cyber-attacks. Moubayed \textit{et al.} ensemble-based anomaly detection system (ADS) \cite{moubayed2021b, moubayed2020a} propose an ensemble feature engineering and optimized random forest-based intrusion detection model to detect Botnet attacks and DNS typo-squatting. Yang \textit{et al.} \cite{yang2021a, yang2021b} propose an adaptive data analytics framework for anomaly detection in evolving and dynamic networks. It analyzes the time-based features in evolving networks and achieves high accuracy of more than 99\% on public datasets. However, its high run-time complexity remains an issue. Injadat \textit{et al.} \cite{injadat2020a, injadat2020b} propose Bayesian optimization-based ML models for network intrusion detections. The proposed models can effectively detect various types of cyberattacks, but they did not analyze the attacking details. Although the above methods achieved high performance on benchmark datasets for anomaly detection, they are designed only to detect the occurrence of cyber-attacks instead of fingerprinting the detailed information of cyber-attacks and victims. It is crucial to obtain the attacking details, as they can help restore the compromised devices, hold the attackers accountable, and prevent future attacks. Forensic analytics techniques can be used to fingerprint the details of network anomalies and attacks.
\subsection{Forensic Data Analytics}
Digital forensic analytics techniques have been applied to several network anomaly detection and cyber-security-related problems. Amato \textit{et al.} \cite{amato2020a} propose a semantic-based method for digital forensic analysis in cyber-security problems. This method enables the generation and retrieval of useful data for digital evidence collection. Koroniotis \textit{et al.} \cite{koroniotis2020a} propose a Deep Neural Network (DNN) and Particle Swarm Optimization (PSO) based forensic architecture to detect botnet attacks. The proposed framework is able to trace the behaviors of cyber-attack events with better performance than other compared methods. Khan \textit{et al.} \cite{khan2019a} present a multi-agent method to conduct digital forensic analysis in storage networks. The access logs collected in a server can be aggregated by the agents for further verification to detect malicious events and cyber-attacks. Although there are many existing forensic techniques, the majority of them are designed for labeled datasets, which is usually difficult to be acquired in real-world network applications. Thus, it is still critical to upgrade existing methods and develop new methodologies to deal with digital forensic analytics problems more effectively \cite{hou2020a}.
\subsection{Service Targeting Attack Detection}
As described in Section \ref{STA}, cyber-attackers usually aim to take down or disrupt services in evolving networks, named \textit{service targeting attacks}. As this chapter focuses on detecting service targeting attacks (\textit{e.g.}, CPAs and DoS attacks) due to their destructiveness, a review of existing research works for service targeting attack detection is conducted. Vasseur \textit{et al.} \cite{vasseur2019a} propose a network anomaly detector that utilizes unsupervised learning algorithms to generate a set of rules from captured network data, thus to train a supervised learning classifier to identify anomalies. However, a convincing validation process is required to verify the generated rules. Conti \textit{et al.} \cite{conti2013a} performed studies to demonstrate that CPAs constitute a serious danger to network security and suggested a lightweight detection approach for reliably detecting CPAs.  Baradaran \textit{et al.} \cite{baradaran2019a} propose an anomalous network traffic detection method that compares the extracted feature values with the predetermined threshold values to determine anomalies and identify DoS attacks. However, determining an appropriate threshold is often difficult. Yao \textit{et al.} \cite{yao2020a} propose a detection scheme by using the grey forecast to predict the popularity of each cached content, and use the estimated popularity information to identify cache pollution attacks. Kumar \textit{et al.} \cite{kumar2019a, moubayed2019a} propose a security framework named Software-Defined Perimeter (SDP) to protect modern networks from DoS attacks. Karami \textit{et al.} \cite{karami2015a} propose an anomaly detection model that uses k-means and PSO algorithms to detect DoS attacks in content-centric networks. However, this method is not evaluated on a representative network dataset. The SDP framework can prevent and defend against upcoming DoS attacks but cannot investigate the past attacks that have already breached the networks. 
\subsection{Cybercrime Related Entity Detection}
Network entities are the physical pieces and components comprising the network, such as client IPs, service nodes, contents, offerings, etc. Detecting network entities affected by cyber-attacks is critical in the anomaly detection process because they can help analyze the attacking details. Detected malicious client IPs are the potential attacker IPs, which can help us locate the attackers. Detecting compromised nodes enables us to locate the target devices and repair network failures. Doctor \textit{et al.} \cite{doctor2019a} propose a system for identifying and mitigating malicious threats on a network by collecting and analyzing network traffic data associated with IP addresses. Malicious IP addresses can be detected, but this system does not consider service provider perspectives. Ayyagari \textit{et al.} \cite{ayyagari2015a} proposed a context-aware network threat detection method that monitors the nodes associated with users to generate a behavior profile for each user. Anomalies can then be detected by comparing each user’s behavior profile with the baseline behavior profile. Qing \textit{et al.} \cite{qing2019a} proposed a network anomaly detection method based on gradient boosting decision tree (GBDT) that extracts node status and routing information features, such as cache hit rate and caching life rate to detect cache pollution attacks, but did not consider other network entities, like client IPs. Pandey \textit{et al.} \cite{pandey2016a} propose an intrusion detection method that uses the agents to identify compromised nodes in wireless sensor networks (WSNs) based on their behavior, which shows good efficiency in small networks. However, root cause analysis was not conducted in the above techniques, which is significant for future attack prevention.
\subsection{Research Gaps}

In summary, in the context of security detective controls to be provided for evolving networks, we identify the following issues with existing methods: 
\begin{enumerate}
\item The complexity of evolving networks makes it difficult to merely identify and combine individual sources of events as reliable indicators of service targeting attacks. For example, distributed DoS (DDoS) attacks are launched by a large number of malicious clients instead of an individual client, so the behaviors of all these malicious clients should be analyzed together to effectively identify DDoS attacks;
\item Existing security technologies fail to model the evolving networks in a way to properly capture data-driven semantics out of “multi-perspective”(s), a concept we consider essential;
\item In evolving networks, the following requirements are not properly fulfilled by existing solutions:
\begin{enumerate}
\item Perspectives are defined in a way to allow their interaction;
\item It is possible for a security baseline to be defined from multiple perspectives. Against such baseline, deviations are detected as network attacks; and
\item The indicator of compromises should be effectively validated for a better security posture in evolving networks;
\end{enumerate}
\end{enumerate}

Thus, solutions for more efficient and cost-effective detection of service targeting attacks in such complex ecosystems are left as open problems. The objectives of the proposed solution are to develop the following techniques:
\begin{enumerate}
\item Holistic multi-perspective cache and service-based digital forensic analytics techniques for anomaly detection in delivery evolving networks;
\item Characterization of network anomalies as attacks to improve security posture management in evolving networks;
\item A mechanism to detect service targeting attacks based on selective features engineering involving multiple perspectives: content (\textit{i.e.}, media, configurations, files, gaming), IP space (\textit{i.e.}, clients, IoTs), service nodes, and offerings (\textit{i.e.}, network slices, accounts);
\item An unsupervised anomaly detection mechanism to fingerprint abnormal contents and malicious IPs associated with attack detection, so as to locate attackers;
\item Hyper-parameter optimization methods and mechanisms to build optimized machine learning models for service-based attacks detection;
\item A composite methodology to infer anomaly detection models for implicit identification of attacks targeting services (\textit{i.e.}, cache pollution, denial of service); and
\item Implicit root cause analysis to identify and validate targeted delivery nodes associated with attack detection.
\end{enumerate}

\section{Multi-Perspective as Intelligence for Anomaly Detection}
\label{prop}
This section provides an overview of the proposed solution, which uses multi-perspective as intelligence for anomaly detection in evolving networks. The potential deployment of the proposed system in evolving networks is also discussed in this section.
\subsection{Security Posture Support in Evolving Networks}
In this chapter, we rely on an approach that aims to digest application-layer logs in evolving networks to identify anomalies for the purpose of hardening their security posture. A security posture refers to the overall security status of the software and hardware assets, networks, services, and information \cite{pour2021a}. A perspective in evolving networks means monitoring and analyzing network states based on the behaviors of a single type of network entity, such as the client and service provider. Analyzing the interactions among multiple perspectives enables people to have a broader view of network behaviors than a single perspective, which is beneficial for accurate cyber-attack detection. 

\begin{figure}[!t]
     \centering
     \includegraphics[width=11.5cm]{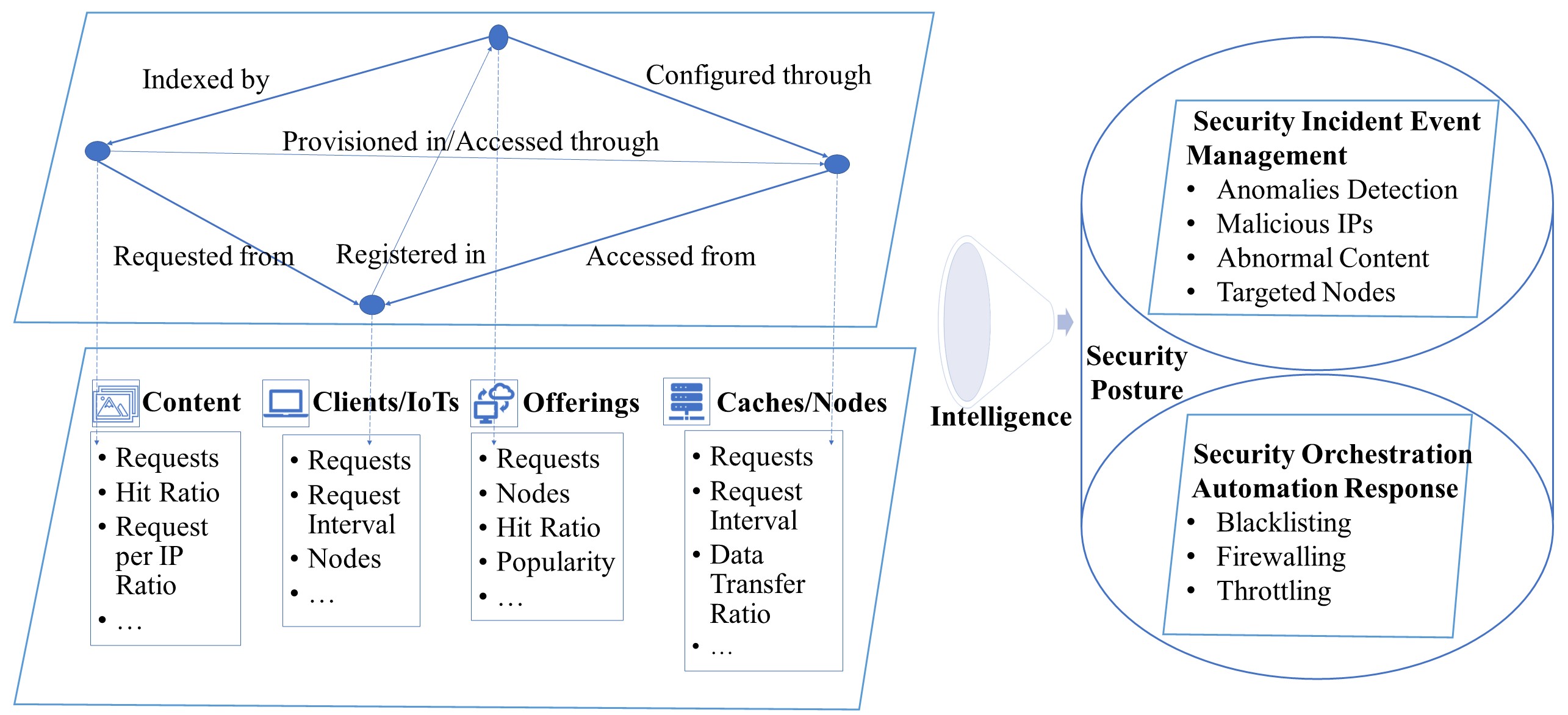}
     \caption{Security posture support for evolving networks. } \label{posture}
\end{figure}

As depicted in Fig. \ref{posture}, knowing the following interactions: 
\begin{enumerate}
\item The service nodes (\textit{i.e.}, proxies, gateways, caches, brokers) located in evolving networks are accessed by clients or IoTs from the IP space;
\item The content (media, files, and configurations) is requested from the clients/IoTs and provisioned or accessed through service nodes;
\item The offerings are indexing the content and configured based on service nodes and caches; and
\item The clients and IoTs are registered in offerings for diverging services.
\end{enumerate}

The solution aims to characterize these perspectives through attributes to infer an intelligence to support the security incident event management (SIEM) and security orchestration, automation, and response (SOAR). By intelligence, we mean the ability to infer a security baseline to identify anomalies like attacking IPs, abnormal content as well as victims (targeted service nodes). The intelligence is used by SOAR to trigger defense and mitigation techniques like dynamic firewalling, throttling, and blacklisting.

\begin{figure}[!t]
     \centering
     \includegraphics[width=11.2cm]{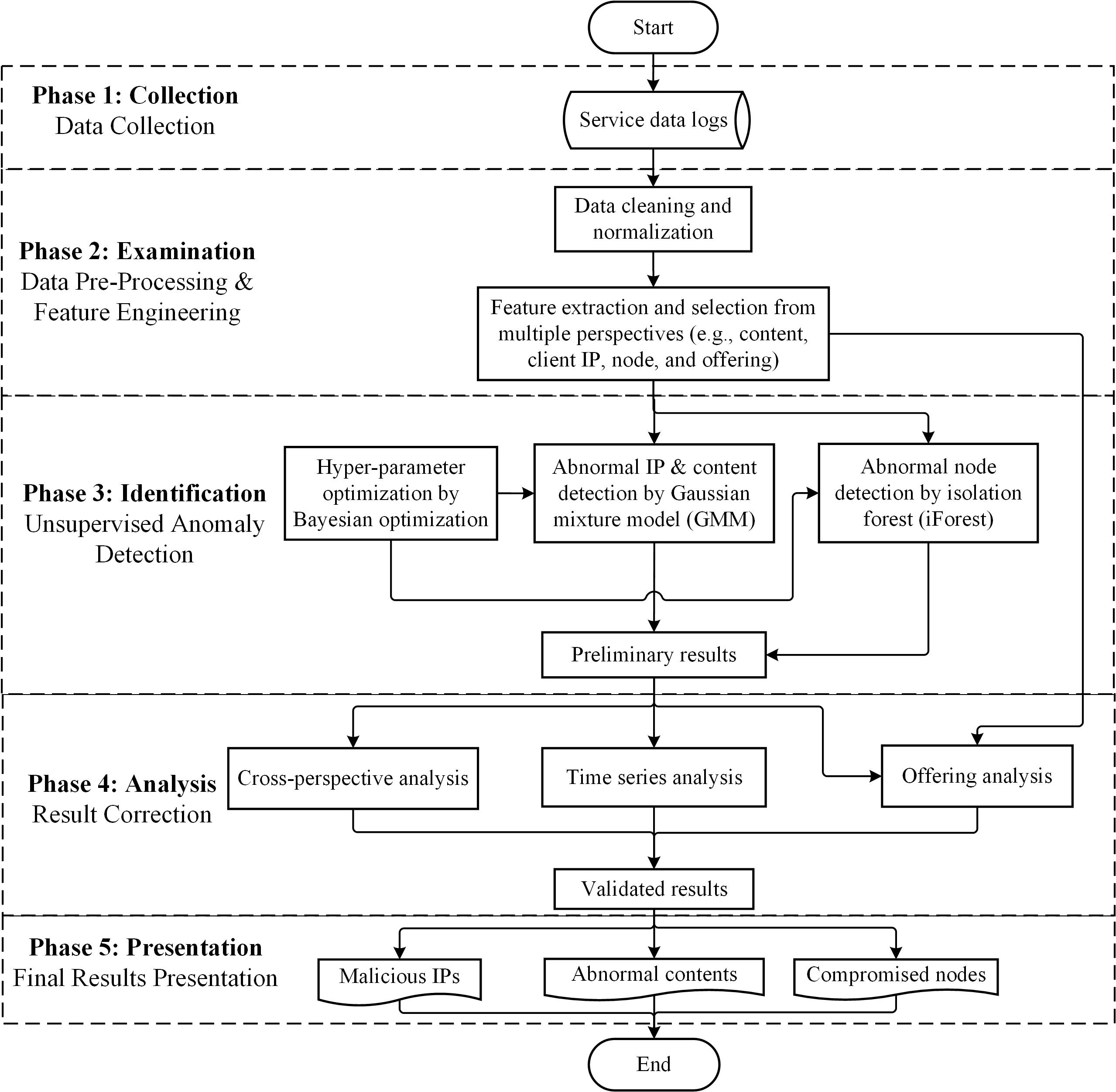}
     \caption{The proposed forensic analytics framework for anomaly detection. } \label{flow}
\end{figure}

\subsection{Digital Forensic Analytics Framework for Anomaly Detection}
\label{overview}
The proposed framework intends to characterize abnormal network events \& attacks and distinguish them from benign events by analyzing network features extracted from multiple perspectives, including client IP, content, node, and offering perspectives. The framework for the proposed anomaly detection model is shown in Fig. \ref{flow}, which consists of five stages: data collection, data pre-processing \& feature engineering, unsupervised anomaly detection, result correction, and final results presentation. These five stages are designed based on the five standard procedures of digital forensic analytics described in Section \ref{DFA}: collection, examination, identification, analysis, and presentation.

In the collection process, access log traces are collected. They consist of web requests received and recorded by multiple service providers in a large evolving network. After collecting the data logs, they are pre-processed in the examination procedure to generate a sanitized version of the data. A multi-perspective feature engineering method is also proposed and utilized in the examination process to extract features and generate new datasets from different perspectives that can accurately depict the behaviors of cyber-attacks from each perspective. In the identification process, the datasets extracted from different perspectives are processed by the proposed unsupervised anomaly detection system, consisting of the Gaussian mixture model (GMM), Isolation forest (iForest), and Bayesian optimization (BO) models to effectively discriminate between abnormal and benign patterns. In the analysis process, the preliminary anomaly detection results will be validated by the proposed comprehensive result correction frameworks to determine the real cyber-attacks and identify the related network entities, such as nodes, contents, and IPs. Lastly, the anomaly detection results, including the detailed information of cyber-attacks types, victims, attackers, and utilized services, will be returned and displayed in the presentation procedure.

\subsection{System Deployment}
\begin{figure}[!t]
     \centering
     \includegraphics[width=11cm]{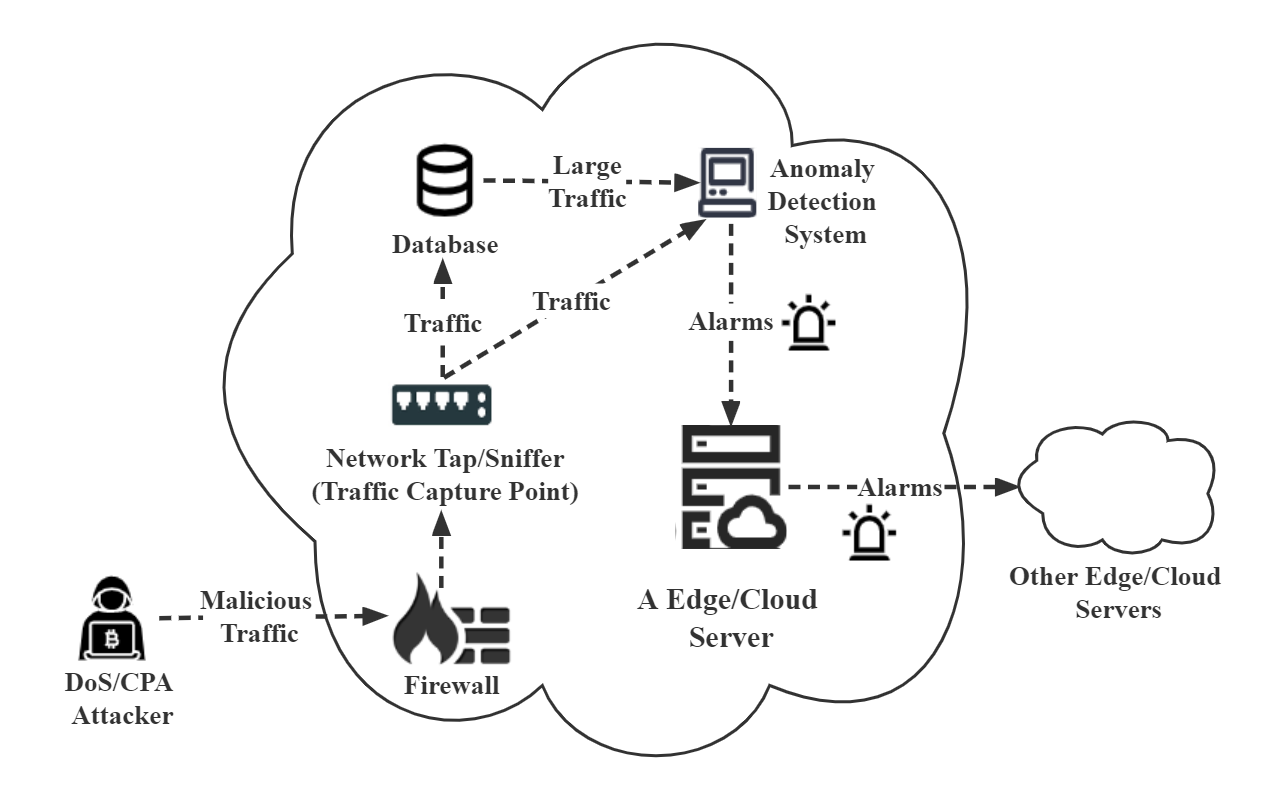}
     \caption{The deployment of the proposed anomaly detection system. } \label{deploy}
\end{figure}

For applying the proposed anomaly detection system to evolving networks, it can be placed in both edge and cloud servers in evolving networks, as shown in Fig. \ref{deploy}.

The proposed system can continuously monitor network traffic in edge servers for abnormal network entity detection and provide warnings to the central server as soon as an attack is launched; thus, the central server can notify other edge servers and take appropriate countermeasures. When installed on the central server, the proposed system can provide a complete picture of the whole network's functioning and can safeguard the central server when attackers use particular edge servers to compromise the central server. Specifically, all network traffic that is not blocked by the first layer of security (\textit{i.e.}, firewalls and authentication) would be recorded and analyzed by the suggested anomaly detection system \cite{yang2022b} on each edge or cloud server. Additionally, the vast volume of traffic may be saved in a database for a thorough examination by the suggested anomaly detection system. If an attack is detected on one of the edge or cloud servers, an alert is raised on all of the edge and cloud servers. As a result, network administrators at the central server and edge servers may take appropriate actions to thwart existing and future assaults.

\section{Data Pre-Processing \& Feature Engineering}
\label{pre}
This section presents the two critical procedures of the proposed forensic analytics anomaly detection framework: data pre-processing and feature engineering. Data pre-processing is the process of improving data quality for more accurate analytics, while feature engineering is the process of extracting or selecting appropriate network features that can better reflect cyber-attack patterns. 

\subsection{Data Collection and Description}
Data collection refers to the initial step of forensic analytics. In our solution, we focus on anomaly detection in the application layer (\textit{e.g.}, HTTP, CoAP, MQTT, LwM2M, industrial protocols) access/event logs. These logs record the information of contextual requests used to run a business use case. For the sake of illustration, clients request content from web servers. The content includes HTML text files, embedded images, videos, audios, and other associated files provided by a web service. Access/Event logs help to identify states and behaviors triggered by contextual applications in a network on a timely basis. A typical sample of general access log data sample is shown below:

\textit{64.0.0.1 - - [11/Dec/2016:05:33:28 -4000] “GET /support.html HTTP/1.1” 200 15340 - “Mozilla/5.0(compatible;Googlebot/2.1;+http:// www.google.com/bot.html)” }

Numerous network fields/features can be collected for the purpose of anomaly detection. Specific features should be chosen for the detection of certain attacks. The major raw features that may be helpful for the detection of service targeting attacks (\textit{i.e.}, DoS attacks and CPAs) are depicted in Table \ref{raw}.

\begin{table}[]
\tbl{Application protocols raw features}
{\begin{tabular}{|p{8em}|p{26em}|}

\hline
\textbf{Feature}               & \textbf{Description}                                                                                                                                        \\ \hline
IP                                     & Client   IP addresses                                                                                                                                                                                                                                                                                                                                           \\ \hline
Timestamp                              & Start timestamp of each request, in format “[dd/mmm/yyyy:hh:mm:ss -zzzz]"                                                                                                                                                                                                                                                                                     \\ \hline
Protocol method                      & \begin{tabular}[c]{@{}p{26em}@{}}   HTTP request methods, (\textit{i.e.}, GET, POST, etc.)\\       CoAP request methods, (\textit{i.e.}, GET, DELETE, PATCH, etc.)\end{tabular}                                                                                                                                                                                            \\ \hline

Status   code                          & \begin{tabular}[c]{@{}p{26em}@{}}HTTP   status code: 2xx represents a successful response; 3xx represents a redirection;   4xx represents a client error; 5xx represents a server error\\    CoAP   status code: 2xx represents a successful response; 4xx represents a client   error; 5xx represents a server error; 7xx represents signaling codes\end{tabular} \\ \hline
Bytes                                  & Bytes   sent across the network in response to a request                                                                                                                                                                                                                                                                                                        \\ \hline
Delivery   time                        & Duration   of a request from start to finish, including delivery of all content, in   milliseconds                                                                                                                                                                                                                                                              \\ \hline
Device/IoT   agent type                & The   type of device/IoT used to make a request, \textit{e.g.}, iPhone, iPad,   desktop, etc.                                                                                                                                                                                                                                                                            \\ \hline
Service   type                         & The   type of service, \textit{e.g.}, live streaming, static, progressive download,   sensor, etc.                                                                                                                                                                                                                                                                       \\ \hline
Service/   Service cache hit indicator & Service   or cache hit/miss indicator, \textit{e.g.}, hit or miss                                                                                                                                                                                                                                                                                                        \\ \hline
Node                                   & Node   name or ID, representing a service provider or device                                                                                                                                                                                                                                                                                                    \\ \hline
Offering                               & The   network account offering name (\textit{e.g.}, streaming, static web content, dedicated   network slice for IoTs applications), indicating a service accessed from the   IP space                                                                                                                                                                                   \\ \hline
Service   content-canonical path       & The   content part of a canonical path, indicating unique content (URLs for HTTP,   CoAP, MQTT URL IoT Agent, etc.)                                                                                                                                                                                                                                             \\ \hline
Content   type                         & The   type of content in each request, \textit{e.g.}, text, video, image, audio,   configuration files, patches, etc.                                                                                                                                                                                                                                                    \\ \hline
\end{tabular}
}
\label{raw}%
\end{table}

\subsection{Data Normalization}
Data normalization is the process of converting data features to be on a similar scale. It is required when features have varying scales, as larger-scale features are often regarded as more important features than smaller range features in the training process of ML models, resulting in misleading predictions. Min-max normalization is an effective normalization method that can reduce the impact of varying feature scales. In min-max normalization, the normalized value of each feature value, $\tilde{x}$, is denoted by \cite{yang2019a}:

\begin{equation}
\tilde{x}=\frac{x-\min f}{\max f-\min f}
\end{equation}
Where $x$ is the original data value, $\max f$ and $\min f$  are the maximum and minimum values of the feature $f$.

Among the normalization methods, min-max normalization is suitable for anomaly detection problems, as it can retain the outlier values (\textit{i.e.}, extremely large or small values) in datasets to help detect anomalies \cite{yang2022b}. Additionally, it can convert all features to be on the same scale of 0 and 1. Thus, it is selected for the proposed framework.
\subsection{Feature Engineering}

\begin{table}[t]

\tbl{Description of extracted features from the content perspective}
{\begin{tabular}{|p{6em}|p{8.5em}|p{18.5em}|}

\hline
\textbf{Perspective}              & \textbf{Feature }                 & \textbf{Description}                                                                                                                         \\ \hline
 & Number of requests     & The total number of requests for per content                                                                                        \\ \cline{2-3} 
                         & Popularity               & The popularity of per content, represented by the normalized number of IPs that sent requests to per content                        \\ \cline{2-3} 
                        & Cache hit rate         & The number of cache hits divided by the total number of requests for per content                                                    \\ \cline{2-3} 
{Content}                          & Request per IP ratio   & The ratio of the total number of requests sent for per content to the total number of IPs which sent requests for per content       \\ \cline{2-3} 
                         & Request per node ratio & The ratio of the total number of requests sent for per content to the total number of nodes which received requests for per content \\ \cline{2-3} 
                         & IP dynamicity & The percentage of the IP changes for per content during a period \\ \hline
\end{tabular}
}
\label{f_content}%

\bigskip

\tbl{Description of extracted features from the node perspective}
{\begin{tabular}{|p{6em}|p{8.5em}|p{18.5em}|}

\hline
\textbf{Perspective}              & \textbf{Feature }                 & \textbf{Description}                                                                                                                         \\ \hline
 & Cache   hit rate                   & The number of cache hits divided by the total number   of requests received by per node                                 \\ \cline{2-3} 
                      & Cache   hit rate of legitimate IPs & Average cache hit rate of IPs which only requested   for popular contents on per node                                   \\ \cline{2-3} 
                    & Data   transfer rate (MB/s)        & The total bytes returned divided by the total   delivery time for per node                                              \\ \cline{2-3} 
Node                        & Request   error rate               & The percentage of requests with errors (4xx or 5xx status code)   received by per node                                              \\ \cline{2-3} 
                      & Average   request popularity       & Average content popularity of requests received by   per node                                                           \\ \cline{2-3} 
                      & Content dynamicity       & The percentage of the changed contents cached in the cache space during a period                                                           \\ \cline{2-3} 
                      & IP dynamicity       & The percentage of the IP changes for per node during a period                                                           \\ \cline{2-3}                       
                      & Offering   request rate    & The percentage of requests sent through per  offering for per node, \textit{e.g.}, “account1: 80\%, account2: 20\%” \\ \hline
\end{tabular}
}
\label{f_node}%
\end{table}

In the raw data, although there are 13 network features for each request, it is difficult to use this data to identify network anomalies, as cyber-attacks and affected entities usually cannot be directly reflected in single requests. Thus, to perform effective anomaly detection, new datasets that can reflect abnormal network behaviors should be generated through feature engineering. This process corresponds to the second phase of standard digital forensic analytics: examination.

Network feature engineering is the process of extracting or selecting appropriate features from the log traces collected from evolving networks’ deployment. To accomplish this, we propose a multi-perspective feature engineering framework to obtain dedicated datasets. It aims to extract network features from four different perspectives: content, service provider (node), IP space, and offerings. These perspectives are meant to index data to characterize a multiple views’ baseline for security. The description of the extracted features for each perspective is shown in Tables \ref{f_content}, \ref{f_node}, \ref{f_ip}, and \ref{f_offering}, respectively.

The obtained datasets or aggregates with the features gathered from four perspectives are first processed separately and then analyzed in conjunction to identify abnormal events or malicious cyber-attacks. In the proposed solution, the intent is to fingerprint malicious IPs and abnormal content that are exploited by attackers, as well as identify compromised service nodes. Intelligence inferred from the offering perspective represents supporting information to validate the abnormality of features’ vectors. 

After feature engineering, the original access log datasets with 452 million requests have been transformed into four datasets from four different perspectives, including the content-based dataset (1.8 million unique contents), node-based dataset (50 unique nodes), client IP-based dataset (1.2 million unique IPs), and offering-based dataset (70 unique offerings).

\begin{table}[t]
\tbl{Description of extracted features from the client IP perspective}
{\begin{tabular}{|p{6em}|p{8.5em}|p{18.5em}|}

\hline
\textbf{Perspective}              & \textbf{Feature }                 & \textbf{Description}                                                                                                                         \\ \hline
 & Number   of requests            & The number of requests sent by per IP                                                                                 \\ \cline{2-3} 
                              & Average   request interval      & The average time interval between consecutive   requests sent by per IP                                                     \\ \cline{2-3} 
                              & Number   of nodes               & The total number of unique nodes that received   requests from per IP                                                       \\ \cline{2-3} 
                              & Number   of contents            & The total number of unique contents requested by per IP                                                                 \\ \cline{2-3} 
Client   IP                              & Request   per content ratio     & The ratio of the total number of requests sent by per   IP to the total number of contents requested by per IP              \\ \cline{2-3} 
                              & Request   per node ratio        & The ratio of the total number of requests sent by per   IP to the total number of nodes which received requests from per IP \\ \cline{2-3} 
                              & Average   request popularity    & Average content popularity of requests sent by per IP                                                                       \\ \cline{2-3} 
                              & Cache   hit rate                & The number of cache hits divided by the total number   of requests sent by per IP                                           \\ \cline{2-3} 
                              & Request   error rate            & The percentage of requests with errors (4xx or 5xx status code) sent by per IP                                                        \\ \cline{2-3} 
                              & Mobile rate            & The percentage of requests sent through mobile devices by per IP                                                        \\ \cline{2-3}              
                              & Offering   request rate & The percentage of requests are sent through per offering for per IP, \textit{e.g.}, “account1: 80\%, account2: 20\%”       \\ \hline
\end{tabular}
}
\label{f_ip}%

\bigskip

\tbl{Description of extracted features from the offering perspective}
{\begin{tabular}{|p{6em}|p{8.5em}|p{18.5em}|}

\hline
\textbf{Perspective}              & \textbf{Feature }                 & \textbf{Description}                                                                                                                         \\ \hline
 & Number   of requests & The number of requests through per offering                                                        \\ \cline{2-3} 
                                         & Number   of nodes    & The total number of unique nodes that received   requests through per offering                           \\ \cline{2-3} 
Offering                                                             & Service   type       & The type of service provided by per offering, \textit{e.g.},   static, live streaming, progressive download, etc. \\ \cline{2-3}                   & Content   type       & The type of content provided by per offering, \textit{e.g.},   text, image, audio, video,  etc.  \\ \cline{2-3}          
                                         & Request   popularity & Average content popularity of requests sent through   per offering                                       \\ 
                                         & Cache   hit rate     & The number of cache hits divided by the total number of requests through per offering \\ \cline{2-3} \hline
\end{tabular}
}
\label{f_offering}%
\end{table}

\subsection{Attack Patterns}
To launch targeting services attacks, attackers typically aim to either disrupt services or alter functionalities. The attacks aiming to disrupt services are known as DoS attacks, while cache pollution is a common type of alteration of service attacks that aims to pollute the nodes’ cache space with unpopular or illegitimate contents in content-based networks.
Based on the potential characteristics of alteration of service and DoS attacks, the extracted features that can directly reflect the attack patterns are selected for each type of attack, as shown in Tables \ref{cpa} and \ref{dos}.

\begin{table}[t]
\tbl{Potential patterns of service alteration/cache pollution attacks}
{\begin{tabular}{|p{5.8em}|p{6em}|p{14.3em}|p{5.5em}|}

\hline
\textbf{Attack Type}           & \textbf{Perspective}                  & \textbf{Feature}                            & \textbf{Abnormal Patterns (Intensity)}                                                      \\ \hline
 &        & Cache   hit rate                   & Low                                                                     \\ \cline{3-4} 
                      &     Node                         & Cache   hit rate of legitimate IPs & Low                                                                     \\ \cline{3-4} 
                      &                              & Data   transfer rate (MB/s)        & Low                                                                     \\ \cline{3-4} 
                      &                              & Average   request popularity       & Low                                                                     \\ \cline{2-4} 
                      & & Number   of requests               & Large                                                                   \\ \cline{3-4} 
Alteration of                     &                              & Average   request interval         & Short                                                                   \\ \cline{3-4} 
 Service                        &   Client   IP                           & Number   of nodes                  & Small                                                                   \\ \cline{3-4} 
       (\textit{e.g.}, CPAs)              &                              & Number of contents        & Large/Small \\ \cline{3-4} 
                      &                              & Average   request popularity       & Low                                                                     \\ \cline{2-4} 
                      &     & Popularity                         & Low                                                                     \\ \cline{3-4} 
                      &     Content                         & Request   per IP ratio             & High/Low      \\ \cline{3-4} 
                      &                              & Request   per node ratio           & High                                                                    \\ \cline{2-4} 
                      & Offering             & Request   popularity               & Low                                                                     \\ \hline
\end{tabular}
}
\label{cpa}%

\bigskip

\tbl{Potential patterns of DoS attacks}
{\begin{tabular}{|p{5.8em}|p{6em}|p{14.3em}|p{5.5em}|}

\hline
\textbf{Attack Type}           & \textbf{Perspective}                  & \textbf{Feature}                            & \textbf{Abnormal Patterns (Intensity)}                                                      \\ \hline
 &         & Cache   hit rate                   & Low               \\ \cline{3-4} 
                      &    Node                          & Cache   hit rate of legitimate IPs & Low               \\ \cline{3-4} 
                      &                              & Data   transfer rate (MB/s)        & Low               \\ \cline{3-4} 
                      &                              & Request   error rate               & High              \\ \cline{2-4} 
                      &  & Number   of requests               & Large             \\ \cline{3-4} 
Denial of                       &                              & Average   request interval         & Short             \\ \cline{3-4} 
        Service              &     Client   IP                         & Number   of nodes                  & Small             \\ \cline{3-4} 
                      &                              & Cache   hit rate                   & Low               \\ \cline{3-4} 
                      &                              & Request   error rate               & High              \\ \cline{2-4} 
                      & Offering             & Cache   hit rate                   & Low               \\ \hline
\end{tabular}
}
\label{dos}%
\end{table}

DoS attacks are launched mainly to exhaust the network resources of serving nodes by sending a sudden burst of requests. Thus, the cache hit rate of the compromised nodes and of their serviced legitimate IPs will be reduced. The data transfer rate will also be decreased due to reduced resources. Many illegitimate requests may also be sent, resulting in an increased request error rate. DoS attacks may be launched by certain IPs that send a sudden burst of requests to certain target nodes.

Similarly, the functionality of compromised nodes will be altered after an alteration of service attacks. The service provided by the affected nodes will also be degraded, causing a reduced cache hit rate and data transfer rate. Moreover, taking CPAs as an example of alteration of service attacks, the compromised nodes’ cache space will be occupied by unpopular contents, since the malicious IPs of potential attackers will send many requests for low-popularity contents to certain target nodes.

\section{Unsupervised Anomaly Detection}
\label{unsu}
\begin{figure}[!t]
     \centering
     \includegraphics[width=11.5cm]{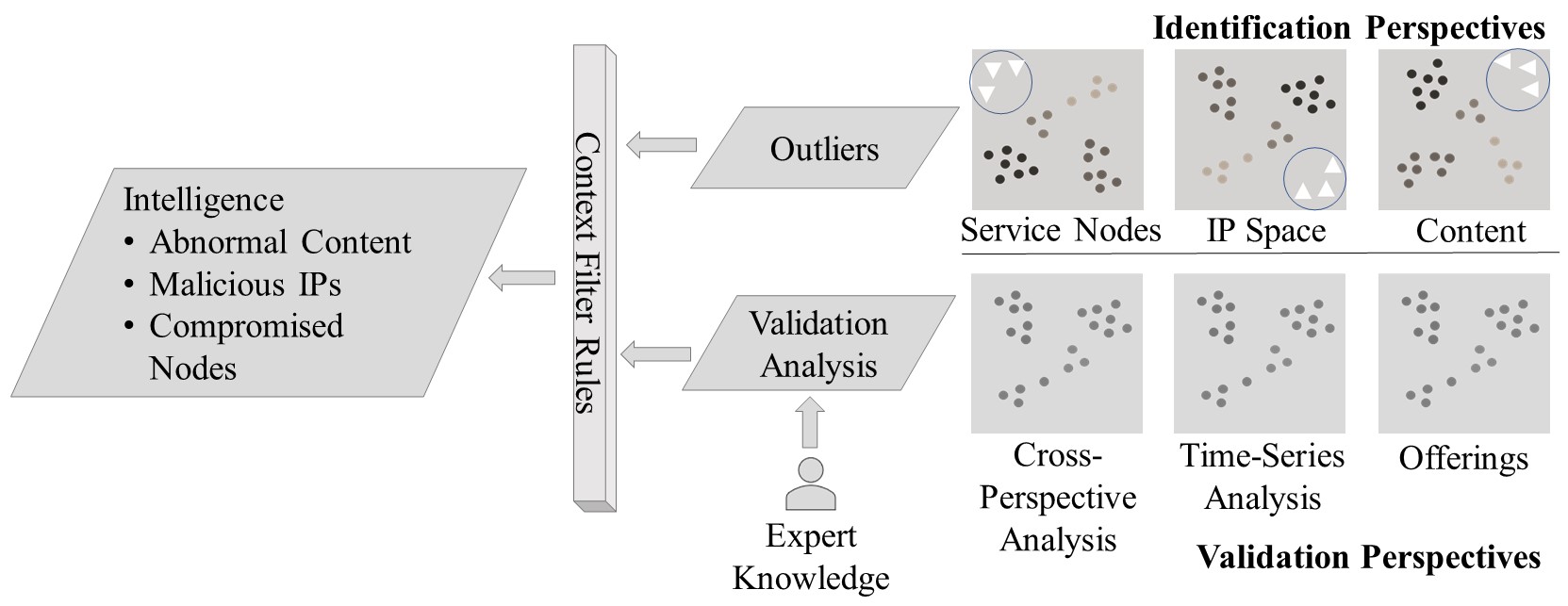}
     \caption{The process of multi-perspective identification and validation of attacks. } \label{multi}
\end{figure}

The proposed unsupervised anomaly detection model is developed based on multiple perspectives, as shown in Fig. \ref{multi}. Unsupervised ML algorithms are first constructed to preliminarily identify numerical anomalies from the identification perspectives, including the service nodes, IP space, and content perspectives. This process corresponds to the third phase of standard digital forensic analytics: identification.

At the next stage, three validation perspectives are considered to validate and correct the anomaly detection results, including cross-perspective, time-series, and offering analysis. This procedure will be discussed in Section \ref{vali}. 

\subsection{Malicious IPs and Content Fingerprinting}
In evolving networks, a node often provides various services or contents for many clients. To separate abnormal client IPs and contents from numerous legitimate IPs and contents, clustering algorithms are promising solutions. Clustering algorithms are a set of unsupervised machine learning models that are used to group unlabeled data samples into multiple clusters \cite{yang2022b}. 

A general process of abnormal IP and content detection consists of the following two steps: 
\begin{enumerate}
\item Use a clustering algorithm to group the client IPs/contents into a sufficient number of clusters; and 
\item Analyze the behaviors of each cluster, and label the IPs/contents in this cluster with “normal” or “abnormal” based on the summarized patterns/characteristics of different types of attacks. 
\end{enumerate}

Gaussian mixture model (GMM) is selected as the clustering algorithm for abnormal content and client IP detection in the proposed solution. GMM uses multiple Gaussian distributions as components to model data points. In GMM, each Gaussian component can be described by a multivariate Gaussian distribution \cite{science2017a}: 

\begin{equation}G(\mathbf{x} \mid \mu, \Sigma)=\frac{1}{(2 \pi)^{\frac{D}{2}}|\Sigma|^{\frac{1}{2}}} e^{-\frac{1}{2}(x-\mu)^{T} \Sigma^{-1}(x-\mu)}\end{equation}

where $x$ is the data sample, $D$ is the number of features, $\mu$ and $\Sigma$ are the mean and covariance of the Gaussian distribution.

A GMM model with $K$ Gaussian distribution components uses a probability density function to estimate the data:
\begin{equation}p(\mathbf{x} \mid \theta)=\sum_{i=1}^{K} \pi_{i} G\left(\mathbf{x} \mid \mu_{i}, \Sigma_{i}\right)\end{equation}
were $\pi_{i}$ is the weight of each Gaussian distribution.

The time complexity of GMM is $O(NKD^2)$ and $O(NKD)$ for $N$ data samples, $K$ Gaussian distributions, and $D$ dimensionalities \cite{pinto2017a}. 

The main reasons for choosing GMM for abnormal content and IP detection are as follows:
\begin{enumerate}
\item 	GMM uses multiple Gaussian distributions to construct a model, which is able to model datasets with complex data shapes or distributions. Thus, it has better adaptability to model real-world network data than many other clustering algorithms, like k-means, which is only effective for globular shape data; 
\item 	Unlike many clustering algorithms (\textit{e.g.}, k-means and hierarchical clustering) that can only return a predicted label for each data sample, GMM can also generate a confidence value that helps us find uncertain data samples to reduce errors; and
\item 	Although GMM has a training time complexity of $O(NKD^2)$ that may be high for high-dimensional datasets, it has a linear run-time complexity of $O(NKD)$ for high run-time efficiency. Additionally, the dimensionality of the network datasets can be low due to the proposed effective feature engineering.
\end{enumerate}

To construct a more effective model that fits the datasets better, the number of clusters/Gaussian components, as the major hyper-parameter of GMM, is optimized by Bayesian optimization (BO) \cite{yang2020a}. BO is a hyperparameter optimization method that uses the previous evaluation results to efficiently determine the future hyperparameter configuration to evaluate. As the two major components of BO, a surrogate model is constructed to fit all the tested hyperparameter values into the objective function, and an acquisition function is used to locate the future hyperparameter values by exploring both the currently-promising regions and the new regions in the search space. BO is used to construct optimized GMMs for the proposed solution since it often exhibits excellent performance in optimizing a small number of continuous or discrete hyper-parameters to which the number of clusters belongs \cite{yang2020a}.

By training two optimized GMMs for client IP and content perspectives, they are used to identify potential malicious client IPs and abnormal contents. For the content-based dataset extracted after feature engineering, 169 contents are preliminarily detected as potential abnormal contents utilized by attackers to launch CPAs, because these contents have very low popularity and got a large number of requests on several target nodes. For the client IP-based dataset, 310 IPs are preliminarily detected as potential DoS IPs because they have sent a sudden burst of requests to certain target nodes, while 54 IPs are identified as potential CPA IPs because they have sent a large number of requests to unpopular contents and targeted on certain nodes. They are the potential IP addresses of cyber-attackers.

\subsection{Compromised Service Nodes Identification}
Compared with the number of client IPs and contents in a general evolving network, the number of nodes is often significantly less since each node can provide a large number of IPs with numerous contents. Utilizing a clustering algorithm to detect compromised nodes is often ineffective because the number of nodes may be too small to form clusters with a sufficient number of similar data samples. Thus, outlier detection algorithms that do not require a large number of data samples are better choices for abnormal node detection. Outlier detection algorithms aim to fingerprint normal patterns and distinguish outliers from the analyzed normal patterns.

A general abnormal node detection process has two main steps: 
\begin{enumerate}
\item Use an outlier detection algorithm to separate numerically abnormal data samples from normal samples; and 
\item Analyze the behaviors of each numerically abnormal node, and label the nodes that match the summarized patterns/characteristics of different types of attacks. 
\end{enumerate}

Isolation forest (iForest) \cite{sun-a}, an outlier detection algorithm that uses true structures to identify isolated data points, is selected for abnormal node selection. IForest is constructed with multiple binary search trees as isolation trees (iTrees) by splitting the instances according to feature values. The number of splittings required to isolate an abnormal sample (\textit{i.e.}, the tree depth) is often less than normal samples since outliers are in relatively sparse regions while normal instances are in relatively dense regions \cite{antonio2017a}. 

The main reasons for selecting the iForest model for abnormal node detection are as follows \cite{sun-a, antonio2017a}: 
\begin{enumerate}
\item Unlike most clustering algorithms, iForest does not require a large data size to build an effective model since iForest uses the low tree depth of data samples to indicate the outliers. Thus, iForest is suitable for node-based datasets with a small number of unique samples;
\item Based on the assumption that in network data, most data samples are normal, while only a small percentage of them belong to anomalies, the iForest model has the adaptability to fit most real-world network data since it can effectively detect outliers in sparse areas of data; 
\item IForest has a low time complexity of $O(N)$ \cite{liu2008a}; and
\item IForest has good interpretability because it uses tree structures to model data.
\end{enumerate}

Similar to GMM, iForest has an important hyperparameter, the contamination level that indicates the percentage of abnormal data samples in the original data. Since it is a continuous hyperparameter, it is also optimized by BO, which shows great effectiveness in optimizing both discrete and continuous variables. Thus, an optimized iForest model is trained for effective abnormal node detection. 

By applying the optimized iForest model to the node-based dataset, 11 potential compromised nodes have been preliminarily detected. These potentially affected nodes have a very low cache hit rate, especially for those legitimate IPs which have sent requests to these nodes. Their data transfer rate and request success rate are also much lower than other normal nodes.

\section{Anomaly Detection Result Correction}
\label{vali}
Although machine learning algorithms, including GMM and iForest, are able to identify numerically abnormal network entities (\textit{e.g.}, content, nodes, and IPs), many prediction errors might occur because many numerical outliers are not true attacks. Certain benign network events, like crowd events and misconfigurations, have similar behaviors as service targeting attacks and can be misclassified as attacks. Hence, a comprehensive validation process should be conducted to reduce the detection errors and distinguish the real service targeting attacks. In the proposed solution, the detected anomalies are corrected from three validation perspectives: cross-perspective, time-series, and offering analysis. Through this multi-perspective framework, the real anomalies affected by service targeting attacks can be effectively identified. In the standard digital forensic analytics process, this validation step corresponds to the fourth phase: analysis.

\begin{figure}[ht]
     \centering
     \includegraphics[width=11.8cm]{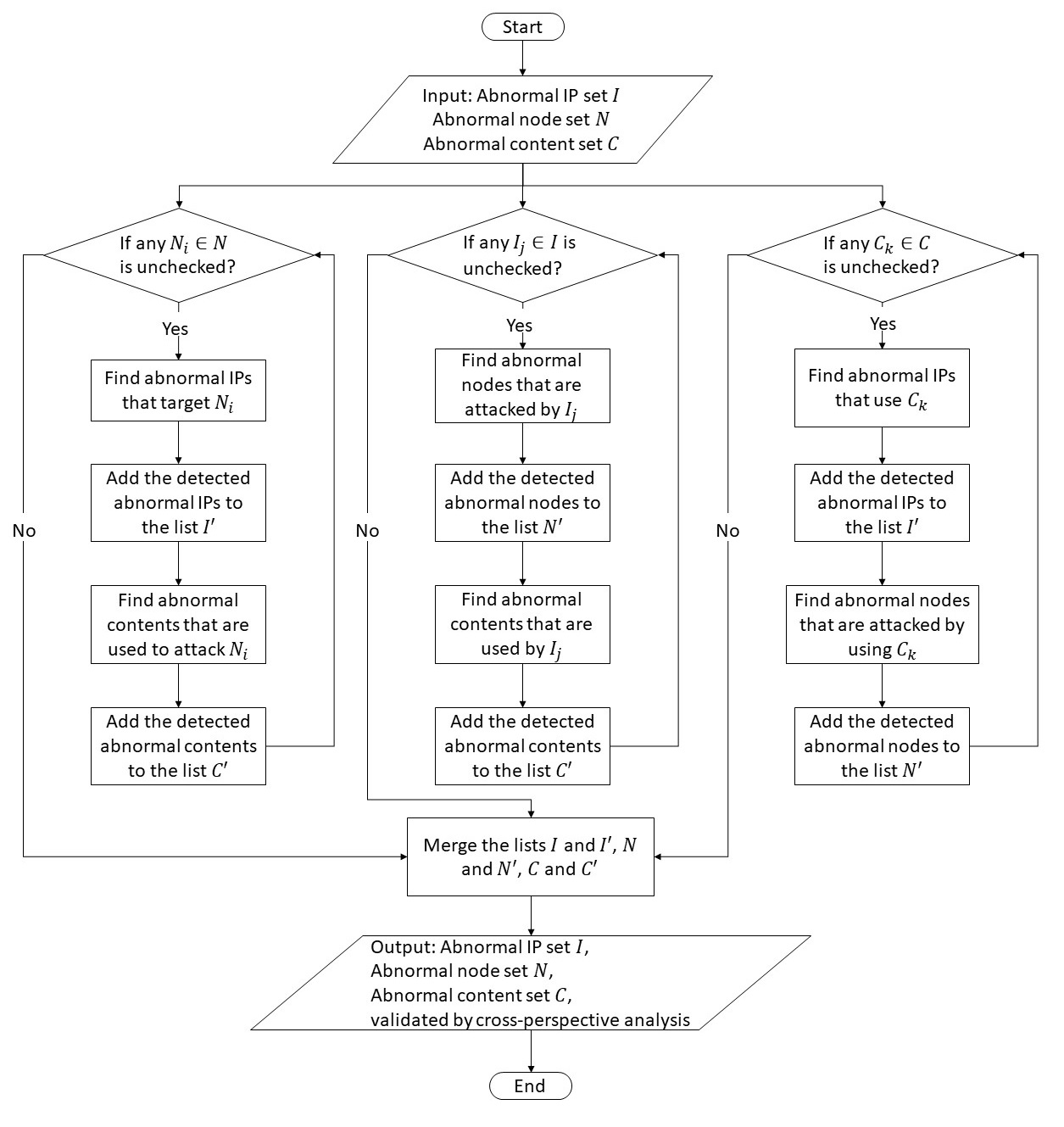}
     \caption{The flow chart of cross-perspective analysis. } \label{cross}
\end{figure}

\subsection{Cross-Perspective Analysis}
After obtaining abnormal detection results from three single perspectives (\textit{i.e.}, content, node, and client IP perspectives), cross-perspective analysis, as shown in Fig. \ref{cross}, is conducted to validate whether the anomalies from different perspectives have correlations. To conduct the cross-perspective analysis, the numerical anomalies detected from each perspective are used to validate the detection results from other perspectives. For instance, the detected affected nodes can be used to validate the potential attacker IPs that have attacked these nodes and the abnormal contents which have been used to pollute these nodes. Additionally, the abnormal contents and IPs detected from their perspectives can also be used to validate the compromised nodes that are targeted by the cyber-attackers who exploited these abnormal IPs and contents to launch attacks. After cross-perspective analysis, the missed true abnormal entities that are affected by service targeting attacks and the false alarms that do not harm any nodes can be identified for more accurate anomaly detection.

\subsection{Time-Series Analysis}

\begin{figure}[htp]
     \centering
     \includegraphics[width=7cm]{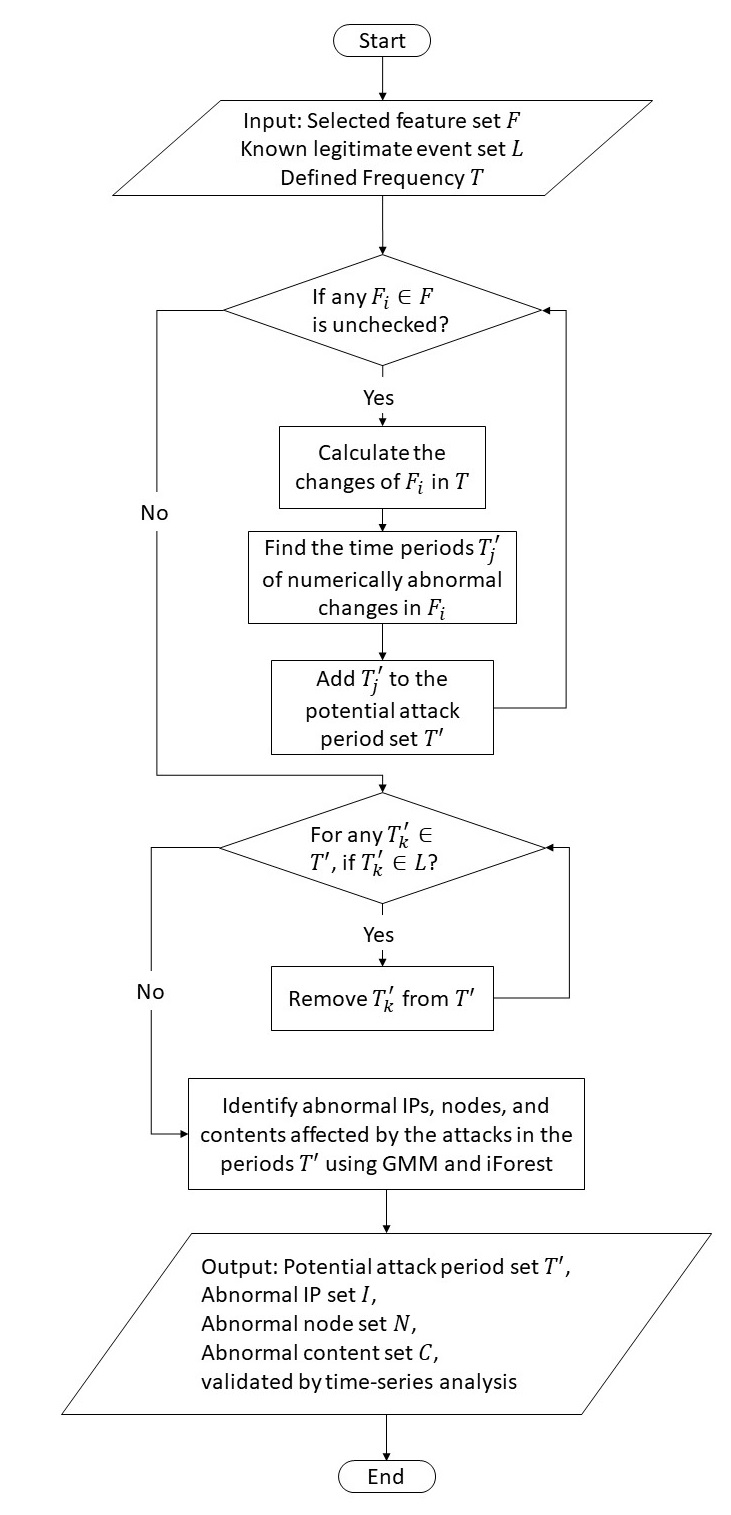}
     \caption{The flow chart of time-series analysis.} \label{series}
\end{figure}

\begin{figure}[htp]
     \centering
     \includegraphics[width=7cm]{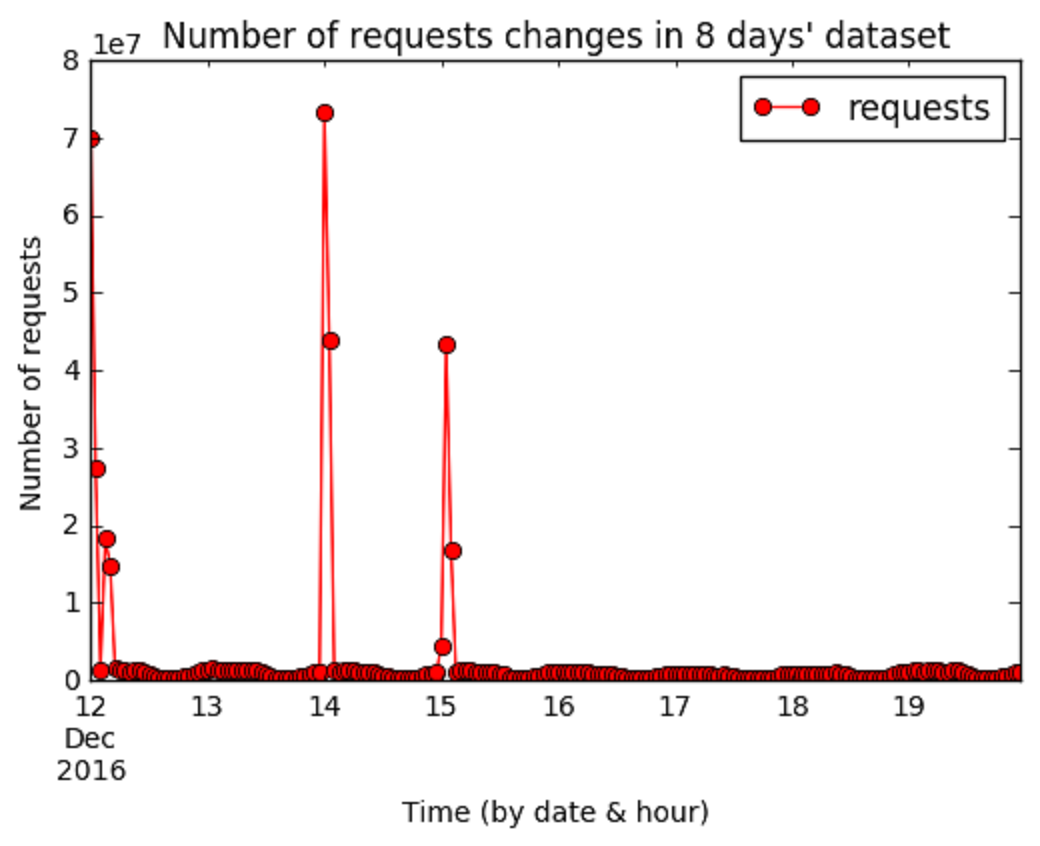}
     \caption{The changes in the number of requests in the original dataset used in this work.} \label{request}
\end{figure}

\begin{figure}[htp]
     \centering
     \includegraphics[width=7cm]{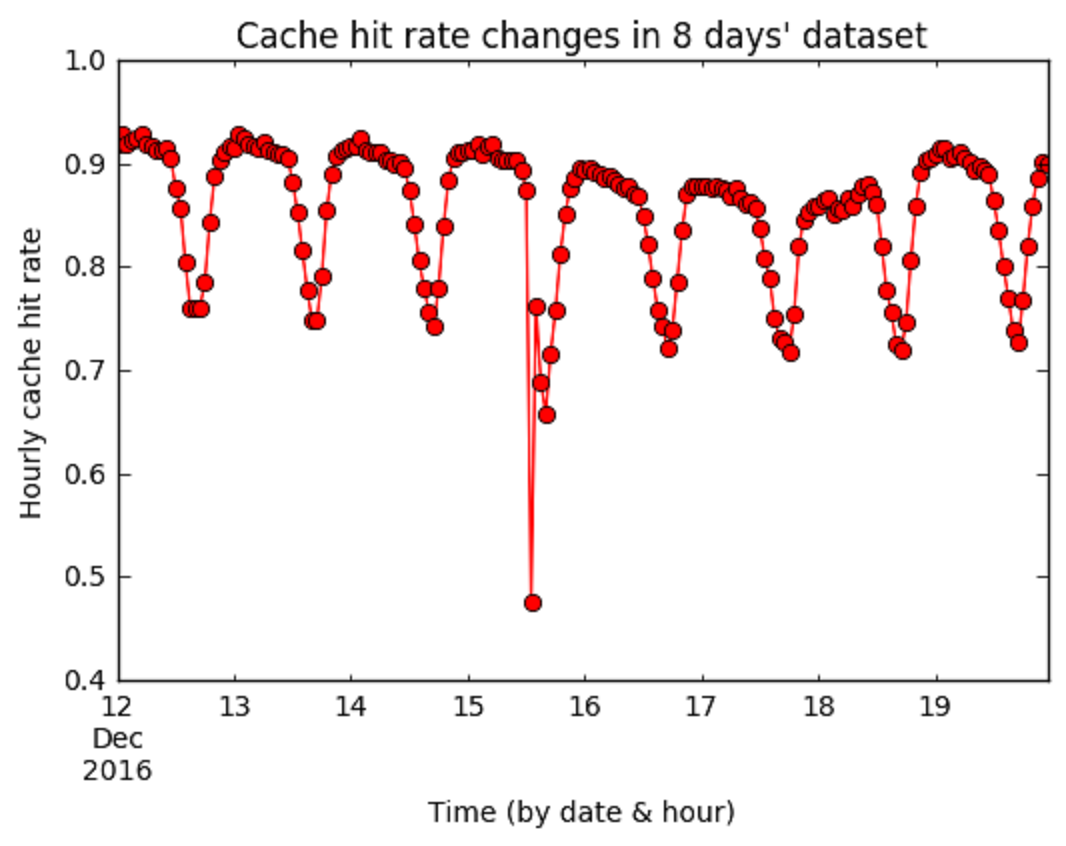}
     \caption{The changes in the cache hit rate in the original dataset used in this work.} \label{cache}
\end{figure}

As the process shown in Fig. \ref{series}, time series analysis is used by examining the changes in particular features over time (\textit{e.g.}, hourly and daily changes) to identify abnormal events and cyber-attacks for result validation. For example, changes in the number of requests over time can be used to identify potential DoS attacks and crowd events when a sudden burst of requests have been sent during certain periods; changes in the cache hit rate over time can be used to identify potential DoS attacks when a sudden burst of error requests occurred to exhaust resources; the changes in the request popularity over time can be used to validate potential CPAs when malicious IPs have sent a large number of requests for low-popularity contents, or compromised nodes have got a large number of requests for low-popularity contents. Time series analysis can help us locate the time periods of potential service targeting attacks and benign events like crowd events by analyzing the network entities affected in these periods.

Take the access log data used in the proposed solution as an example, the changes in the hourly number of requests and hourly cache hit rate in the original datasets are shown in Figs. \ref{request} and \ref{cache}. From Fig. \ref{request}, it is noticeable that on Days 1, 3, and 4, there are three sudden bursts of requests. They can be potential crowd events or DoS attacks. Similarly, it can be seen in Fig. \ref{cache} that the cache hit rate significantly dropped on Day 4, which could also be a potential DoS attack or CPA. Further analysis can be conducted to determine the detailed information of the detected abnormal events.

\begin{figure}[htp]
     \centering
     \includegraphics[width=6.8cm]{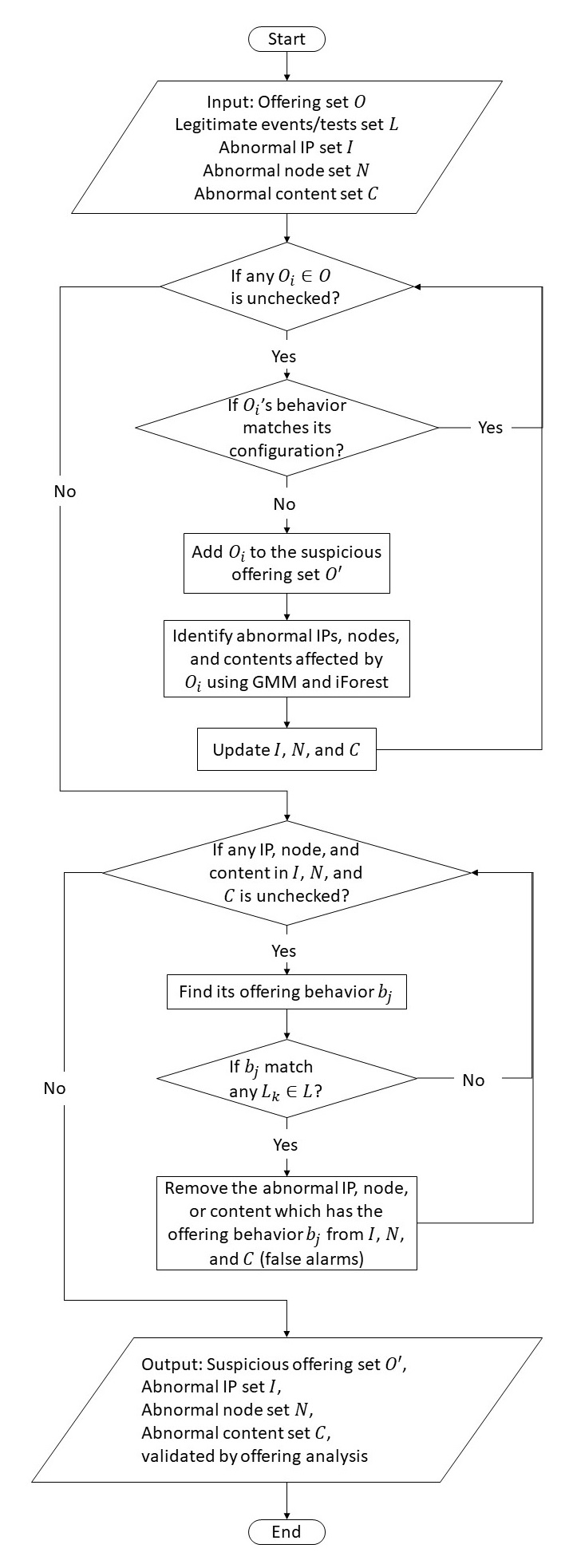}
     \caption{The flow chart of offering analysis. } \label{offering}
\end{figure}

\subsection{Offering Analysis}
Offerings refer to the streaming sources that deliver certain services, like static and live streaming services. Their configurations determine the behaviors of their served IPs. Offering analysis is the process of validating anomaly detection results according to the configurations and behaviors of offerings to separate true cyber-attacks from benign outliers. The general process of offering analysis is shown in Fig. \ref{offering}. Normally, an offering’s potential behaviors can be estimated by its configurations. If service targeting attacks occur, the affected offerings may perform abnormal behaviors, which may assist us in validating the network anomalies serviced by these offerings. For instance, if an offering that is configured to be a static content stream has got a large number of requests for non-existent live streaming content, it might be used by DoS attackers to overwhelm certain target nodes. On the other hand, if an offering is configured for the benign tests of old progressive download videos by network administrators, it may receive numerous requests for unpopular content, which is similar to the behaviors of cache pollution attacks; thus, its affected network entities may be misidentified as false alarms by machine learning models. Offering analysis requires offering configuration information and expert knowledge by comparing offerings’ configurations with their actual behaviors. Through offering analysis, the false alarms obtained from benign events can be reduced for more accurate anomaly detection.

\subsection{Results Summary}
Finally, the real cyber-attacks and their affected network entities have been identified through the proposed forensic analytics process. Specifically, the proposed solution has detected 14 compromised nodes, including 12 nodes attacked by DoS attacks and 2 nodes attacked by CPAs; 155 client IPs have been identified as malicious IPs of potential cyber-attackers, including 122 DoS IPs and 33 CPA IPs; 56 contents have been exploited by attackers to launch CPA attacks. Moreover, among the detected anomalies, 384 false positives and 65 false negatives have been identified and removed using the proposed comprehensive result correction method. All the anomaly detection results have been validated by multiple cybersecurity experts and industrial partner security network engineers.

The detected real anomalies, including the malicious IP addresses, compromised nodes, and abnormal contents, as well as their corresponding attack types, are then presented to trigger corresponding countermeasures, like blacklisting the attacker IPs and abnormal contents, and isolating or recovering the compromised nodes. This phase corresponds to the last step of the standard digital forensic analytics process: presentation.

\section{Summary}
\label{conc}
Most modern networks are evolving networks that are dynamically changed and updated to provide continuous services and functionalities. Due to the increasing number of cyber-threats and crimes, it is crucial to enhance the security of modern networks through digital forensic analytics techniques. Cyber-security and anomaly detection are important applications of digital forensic analytics, which aim to identify data patterns and user behaviors to recognize network anomalies and cyber-attacks. The detailed and comprehensive view of network anomalies enables us to locate attackers, victim devices, and exploited services, so as to convict and stop cybercrimes. In this work, we propose a multi-perspective anomaly detection framework by integrating the five major phases of standard digital forensic analytics: collection, examination, identification, analysis, and presentation. Through the proposed multi-perspective feature engineering, unsupervised anomaly detection, and comprehensive result correction approaches, real-world access log data collected in evolving networks can be effectively processed to fingerprint the service targeting attacks and the affected nodes, malicious attacker IPs, and abnormal contents. The information of these detected anomalies can be used as important digital evidence to solve cybercrimes.

\section*{Acknowledgment}
This chapter is partially supported by the Natural Sciences and Engineering Research Council of Canada (NSERC) [NSERC Strategic Partnership Grant STPGP – 521537] and Ericsson Canada.

Electronic version of an article published as [Book Series: World Scientific Series in Digital Forensics and Cybersecurity, Volume 2, Innovations in Digital Forensics, 2023, Pages 99-137] [DOI: 10.1142/9789811273209\_0004] \textcopyright\ copyright World Scientific Publishing Company [\href{https://doi.org/10.1142/9789811273209_0004}{https://doi.org/10.1142/9789811273209\_0004}]



\end{document}